%% file: 323_RG_Rugg.tex
\documentclass[a4paper]{jpconf}
\usepackage{lmodern} 
\usepackage{amsmath}
\usepackage[T1]{fontenc}
\usepackage[utf8]{inputenc} % input encoding
\usepackage{graphicx}  %to include figures
\usepackage[table,x11names]{xcolor}
\usepackage[pdftex,bookmarksopen,bookmarksnumbered,linktoc=all,hidelinks]{hyperref}


\begin{document}
\title{Interacting dark energy models}

\author{Robert Rugg$^1$, Shambel Sahlu$^{1,2,3}$, and Amare Abebe$^{1,4}$}
\address{$^1$ Centre for Space Research, North-West University, Potchefstroom 2520, South Africa}
\address{$^{2}$Entoto Observatory and Research Centre, Space Science and Geospatial Institute, Addis Ababa Ethiopia}
\address{$^{3}$Department of Physics, Wolkite University, Wolkite, Ethiopia}
\address{$^4$National Institute for Theoretical and Computational Sciences (NITheCS), South Africa}

%only put the email of the main corresponding author, not of all authors.
\ead{\url{31770312@mynwu.ac.za}}

\begin{abstract}
This work focuses on two linear interaction models between dark matter and dark energy, which are proposed as key factors in explaining cosmic history, specifically the late-time accelerating expansion of the universe. Both models are constrained using a Markov chain Monte Carlo analysis (MCMC) using different sets of observational data. The analysis was composed using the Pantheon data set, consisting of 1048 points of SNIa distance moduli measurements from the Pantheon analysis and the Observed Hubble Parameter (OHD) data set using Baryon acoustic Oscillation (BAO), consisting of 57 data points using distance and expansion rate measurement. Both models showed promising results with the OHD data (BAO), with a  interaction that results in a higher dark matter content of $56\%$ and $44\%$, and a Hubble parameter of $65.7 \pm 3 km$ $s^{-1}Mpc^{-1}$  and $65.8 \pm 3 km$ $s^{-1}Mpc^{-1}$  for the interaction dependent on dark matter and dark energy respectively. The pantheon data set however predicted a reverse interaction for both models which does not follow initial assumptions that were made. The pantheon data measured a dark matter content of $18\%$ and $20\%$ with a Hubble parameter of $72.1 \pm 0.003 km$ $s^{-1}Mpc^{-1}$  and $72.3 \pm 0.004 km$ $s^{-1}Mpc^{-1}$. The constrained results are used to revisit the coincidence problem and other problems in standard cosmology. The analysis provided a discrepancy between the different data sets with one having a large error margin.
\end{abstract}

\section{Introduction}
The standard model of cosmology ($\Lambda$CDM) is currently plagued with shortcomings such as dark matter and dark energy concepts which remain in large disagreement, emphasizing the unknown nature of dark energy driving cosmic expansion and dark matter which plays a huge role in large-scale structures of the universe \cite{Bertone_2005,Riess_1998}. Some extra shortcomings include the coincidence problem which describes a narrow time period in which the dark energy and dark matter densities are of the same magnitude. This narrow time period is coincidentally apparent today in which the $\Lambda$CDM model provides no viable answer as to the nature of this occurrence \cite{Coincidence}.  Furthermore, a huge discrepancy of as much as 120 orders of magnitude in the value of the cosmological constant predicted by the theory of General Relativity and Quantum Field theory has left both theories in doubt \cite{Constant}. Another issue that has come to light is the famous Hubble tension, which refers to the discrepancy in the measurement of the Hubble constant ($H_0$) between Cosmic microwave background (CMB) measurements and measurements involving ladder distances. Both early and late-time techniques yield distinct measurements for the Hubble constant of $67.37 \pm 0.5$ and $70.30 \pm 1.04$ in $km/s/Mpc$ for CMB and Supernova Type Ia (SNIa) measurements respectively \cite{Tension}. 
\\ \\
Different approaches to the idea of solving the above-stated issues include modified theories of gravity involving torsion, quintessence involving a phantom field,  braneworld extensions, and a few other theories of a similar nature\cite{Randall_1999,Bahamonde_2015,Bludman_2002}. A more recent approach to addressing these shortcomings involves introducing an interaction between dark energy and dark matter  \cite{Interaction}. $\Lambda$CDM is built on the idea that each sector namely radiation, matter (baryonic matter and dark matter) and dark energy do not interact, however the laws of conservation could include interacting sectors (such as a matter and dark energy interaction), provided that the entirety of these sectors remain conserved \cite{Interaction, Marcel}. The interaction itself could provide a time-dependent cosmological constant which in turn could alleviate the cosmological constant problem \cite{Marsh_2017}. A detailed analysis of the interaction could be paired with ladder distance observational data such as SuperNova type Ia (SNIa), Baryonic Acoustic Oscillation (BAO), and Cosmic Chromometer (CC) data. Furthermore, the model could then be tested against the $\Lambda$CDM model and could lead to changes in the evaluation of the energy densities, alleviating the coincidence problem.
\\ \\
For this work, the organization of this paper is as follows: In section 2, the theoretical background of the interaction is introduced, where two linear models are proposed dependent on the energy densities. Section 3 covers the observational constraints using SNIa and OHD data where the model undergoes a full statistical analysis as to which a best fit result can be made. Further results on how the interaction models (with associated constraints) impact the coincidence problem. Finally, in section 4, conclusions based on the model are expressed to close the paper.

\section{Interacting dark energy}
In the standard model of cosmology, the continuity equation describes conservation in each component of the universe (Matter, radiation, and dark energy). In General Relativity, the continuity equation expresses no interaction between the components of the universe. This is represented by the zero on the right-hand side of equation (1).
The continuity equation is provided below:
\begin{align}
\dot{\rho}_i + 3H(P_i + \rho_i) = 0,  
\end{align}
where the subscript $i$ resembles the specific energy density such as matter (m), dark energy (de), and radiation (r), $\rho$ is the energy density, $P$ the associated pressure, and $H$ the Hubble parameter, $H = \frac{\dot{a}}{a}$ with a being the scale factor. Dark energy and dark matter is often associated with the characteristic of only interacting with gravity. For this reason, any interaction between these two sectors and radiation are neglected. A interaction (Q) can be introduced to the right hand side of the continuity equation such the total conservation equation (i.e. the summation of the continuity equations) remains conserved. In this work, the interaction shall take place between matter as a whole (dark matter and baryonic matter) and dark energy. No conservation laws are breached as the entirety of the interaction remains conserved \cite{Interaction}. Thus, the following continuity equations are proposed:
\begin{align}
&\dot{\rho}_{m} + 3H\rho_{m} = Q,
\\
&\dot{\rho}_{de} + 3H(\rho_{de} + P_{de}) = -Q,
\\
&\dot{\rho}_{r} + 3H(\rho_{r}+P_r) = 0,
\end{align}
where the subscripts $(m)$, $(de)$ and $(r)$ refer to matter, dark energy and radiation respectively \cite{Interaction, Marcel}. Here Q is the term associated to the interaction where the negative sign in front of the Q represents the direction of energy transfer. Thus, for this model dark energy transfers into matter ($DE\rightarrow M$) is the initial estimation. From the right hand side of the continuity equation, it clear that Q is a function of the energy density multiplied by a quantity with units of inverse time \cite{Interaction}. Thus, it makes complete sense to introduce the Hubble parameter such that $Q = Q(H_{\rho_m}, H_{\rho_{DE}})$,
which can then be expanded using taylor expansion as:
\[ Q(H \rho_m, H \rho_{DE}) \approx Q_0 + \frac{\partial Q}{\partial (H \rho_m)}\bigg|_0 (H \rho_m) + \frac{\partial Q}{\partial (H \rho_{DE})}\bigg|_0 (H \rho_{DE}) + \dots
\approx \lambda_m H_{\rho_m} + \lambda_{DE}H_{\rho_{DE}},\]
where $Q_0$ is the value of such that $H\rho_m = H\rho_{DE} = 0$. Thus, two linear models can be drawn up dependant on either of the two interacting densities, such as:
\begin{align*}
Q = 3\eta H\rho_{DE} \text{ and } Q = 3\beta H \rho_m,
\end{align*}
where $\eta$ and $\beta$ represent a dimensionless coupling indicating the magnitude of the interaction. When $\beta = \alpha = 0$, the model reverts back to the $\Lambda$CDM case where no interaction occurs\cite{Interaction}. To distinguish between the two models, the models are named after the coupling terms, hence the $\beta$ and $\eta$ model. The differential equations can be solved analytically through the variation of parameter method \footnote{Using the equation of state in both models such that $w_i = \frac{P_i}{\rho_i}$, where $w_{de} = -1$ and $w_{m} = 0$.}.

\begin{table}[ht]
\centering % centering table
\setlength{\tabcolsep}{0.5em} % for the horizontal padding
{\renewcommand{\arraystretch}{1.2}% for the vertical padding
\begin{tabular}{c c} % creating 10 columns
\hline \hline
\multicolumn{2}{c}{$\beta$ model} \\
\hline 
  $\Omega_{m}$ & $ \Omega_{m,0}(z + 1)^{-3(\beta - 1)}$ \\
   $\Omega_{DE}$ &  $\frac{\beta}{\beta + w} \Omega_{m,0}(z+1)^{(\beta - 1)} + \Omega_{DE,0} - \frac{\beta }{\beta + w}\Omega_{m,0}(z+1)^{3(1+w)}$\\
  
   \hline
  \multicolumn{2}{c}{$\eta$ model} \\
 \hline 
  $\Omega_m$ & $-\frac{\eta}{\eta + w} \Omega_{DE,0} (z+ 1)^{3(w+1+\eta)} +  \Omega_{m,0} + \frac{\eta}{\eta + w} \Omega_{DE,0}(z+1)^{3}$ \\
   $\Omega_{DE}$ & $\Omega_{DE,0}(z+1)^{3(1+w+\eta)}$ \\
\hline 
\end{tabular}}
\caption{\label{tab:mytable}The exact solution of the fractional energy density has been shown using the interaction model. For the case of $\beta = \eta = 0$, the solutions are reduced to the $\Lambda$CDM model. This is expected due to the idea that $\Lambda$CDM is built with no interaction between sectors of the universe. As compared to the $\Lambda$CDM model, a few extra terms have arisen that can be directly linked to the interaction.}
\end{table}
\section{Results and discussions}
\subsection{Data and method} The models were fitted using the Supernova Type Ia data set and a combined data set of Observable Hubble Data (OHD). The supernova dataset, comprising the Pantheon analysis, includes 1048 data points from distinct Type Ia supernovae (SNe Ia). This dataset incorporates both statistical and systematic errors and is additionally distance-calibrated to ensure accurate luminosity distance estimates \cite{Scolnic_2018}. The OHD is a combined data of 26 measurements of the Hubble parameter from the baryon acoustic oscillations (BAO) survey \cite{Farooq_2013}, and 31 data points galaxy surveys known as cosmic chronometers (CC) \cite{shambel}. The data was fitted using Markov Chain Monte Carlo (MCMC) methods with the publicly available emcee python package \cite{emcee, Renier}. No restriction was taken to only consider dark energy decaying into dark matter scenarios. Instead, the entirety of both the negative and positive Q values are taken into account. A generalized log-likelihood was used as:
\begin{equation*}
L = -0.5\left(\frac{(\text{Observerational} - \text{Theoretical})^2}{(\text{Error})^2}\right)
\end{equation*}
where "Observational" refers to the value predicted by the data sets mentioned above with it's corresponding error labeled by "error", and "Theoretical" refers to the value predicted by the theory.
\begin{figure}[!ht]
\centering
\includegraphics[width=0.49\textwidth]{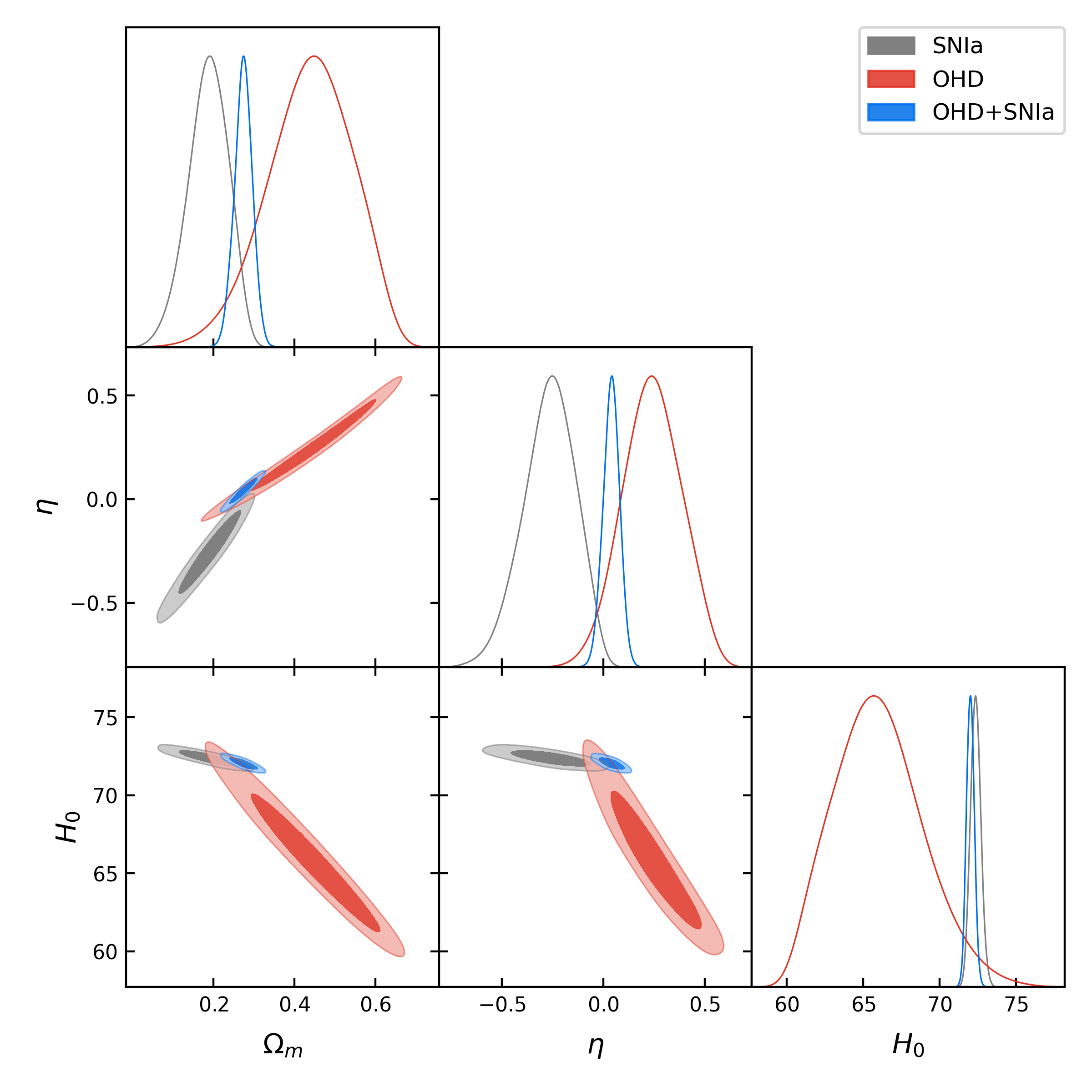}
\includegraphics[width=0.49\textwidth]{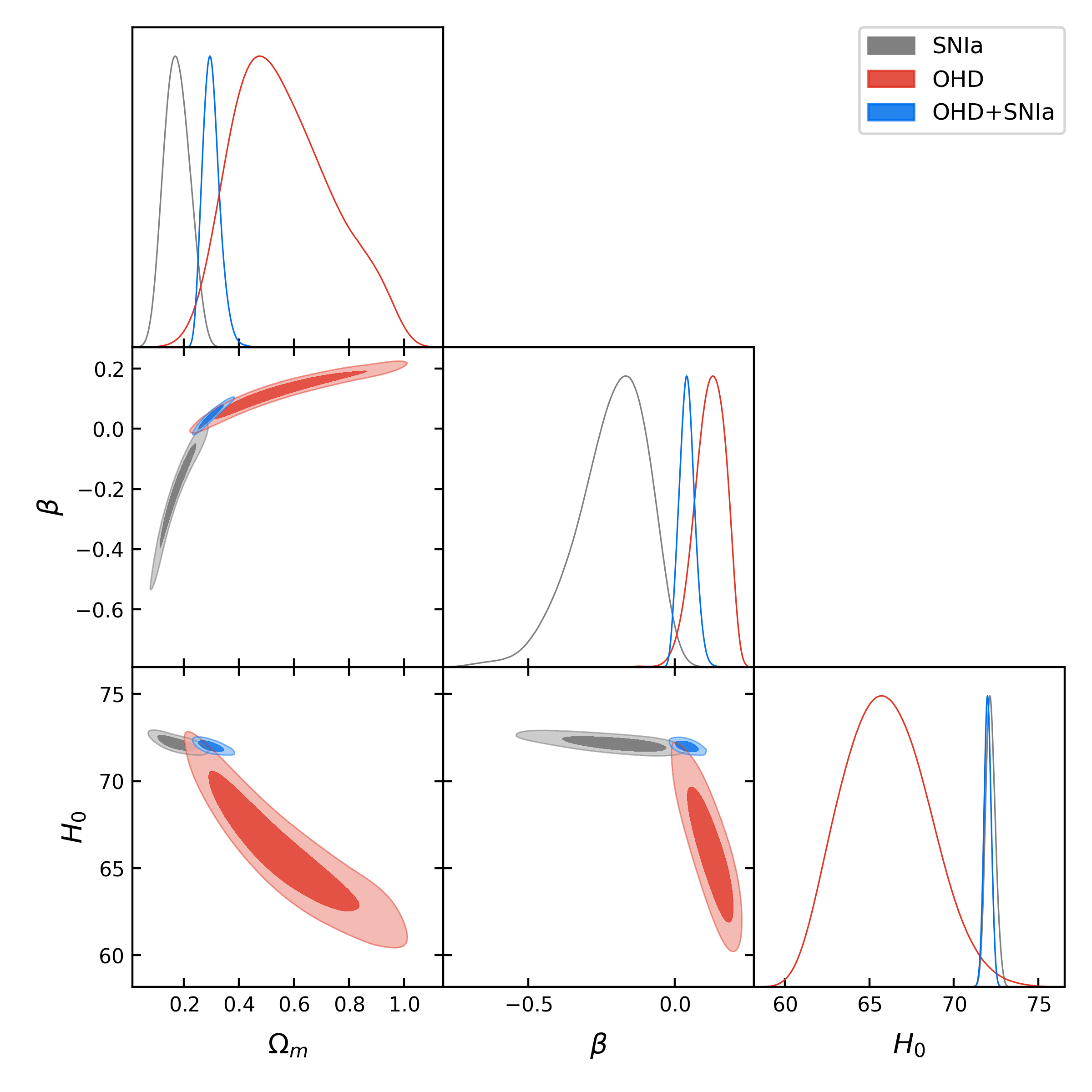}
\caption{\label{fig-sidebyside} (\emph{left}) Best fit model constraints of the $\eta$ model with SNIa, OHD  and a combination of both data sets. (\emph{right}) Best fit model constraints of the $\beta$ model with SNIa, OHD and a combination of both data sets. From this analysis, it can be observed that the $\eta$ model was constrained with regards to the values following in the table below.}
\end{figure}

\begin{table}[ht]
\centering % centering table
\setlength{\tabcolsep}{0.5em} % for the horizontal padding
{\renewcommand{\arraystretch}{1.2}% for the vertical padding
\begin{tabular}{c c c c c c c} % creating 10 columns
\hline\hline 
Model & Data set &$\Omega_m$ & Coupling & $H_0$ $(km/s)/Mpc$ & AIC & BIC\\ 
\hline 
$\Lambda$CDM & SNIa & $0.28 \pm 0.01$ & NA & $71.85 \pm 0.22$ & 1039.6 &  1049.6\\
   $\Lambda$CDM & OHD & $0.25 \pm 0.02$ & NA & $70.79 \pm 1.23$ & 39.3 & 43.1\\
   $\Lambda$CDM & OHD+SNIa & $0.25 \pm 0.02$ & NA & $72.19 \pm 0.18$ & 1082.76 & 1147.32\\
$\beta$ & SNIa & $0.171 \pm 0.054$  &  $-0.203 \pm 0.131$ & $72.158\pm 0.275$ & 1038.7 & 1053.6\\
     & OHD & $0.555 \pm 0.128$  & $0.128 \pm 0.057$  & $65.773\pm 2.617$ & 30.8 & 36.5\\
     & SNIa + OHD & $0.297 \pm 0.023$  &  $0.041 \pm 0.016$ & $72.014\pm 0.045$ & 1129.47 & 1143.96\\
$\eta$ & SNIa & $0.186 \pm 0.055$  & $-0.265 \pm 0.144$  & $72.371\pm 0.355$ & 1038.5 & 1053.6\\
     & OHD & $0.445 \pm 0.115$  &  $0.243 \pm 0.151$ & $65.806 \pm 3.216$  & 31.4 & 37.2\\
     & SNIa + OHD & $0.276 \pm 0.012$  &  $0.042 \pm 0.023$ & $72.024\pm 0.020$ & 1128.96 & 1144.47\\
\hline 
\end{tabular}}
\caption{\label{tab:mytable}The results captured in the table are values provided by the MCMC analysis. $\Omega_m$ represents a dimensionless matter density and $H_0$ the Hubble constant. Only the SNIa results are shown for the $\Lambda$CDM model as it only serves as a bench mark }
\end{table}

\subsection{Observational constraints} An interesting idea appears in the MCMC analysis of both the $\eta$ and $\beta$ model, with OHD data suggesting a DE$\rightarrow$M interaction as initially suggested by the equations above, but the SNIa suggests a M$\rightarrow$DE interaction. Immediately, this creates a tension between the SNIa and OHD data. It must be acknowledged however that the OHD data is accompanied by a large error and thus, accounting for a wider Gaussian distribution which ultimately makes it more accommodating for various models. Further more, the width of the Gaussian distribution for the Hubble constant includes values from $\pm 60$ to $\pm 70$, which is likely due to the high error margin. This is a rather large area of possible values as compared to SNIa which has a narrow peak. Statistically, this shows how SNIa data is better suited to make conclusions of the models. The main result that should be considered is the OHD+SNIa result which can be seen from the graph as a small positive interaction DE$\rightarrow$M.
\\ \\
According to \cite{Elanora}, a similar Hubble constant was recorded using Plank 2018 data. However, the models do differ to a certain aspect in which the model in this paper interacts entirely with matter and not just dark matter as in \cite{Elanora}. However due to the tension in the SNIa and Plank 2018 data, this gives reason to believe that Plank 2018 data could favour the model in this paper which interacts entirely with matter.
\\ \\
Using the Akaike information criterion (AIC) is a refined technique based on in-sample fit to estimate the likelihood of a model and the Bayesian information criterion (BIC) measures the trade-off between model fit and complexity of the model \cite{Stats}. As shown in the table above, the statistical analysis indicates that the $\beta$ and $\eta$ models fit the individual datasets better than the $\Lambda$CDM model. However, for the combined dataset ('SNIa + OHD'), the AIC analysis suggests that $\Lambda$CDM provides a better fit, while the BIC analysis favours the $\beta$ model. Thus, the statistical analysis does not clearly favour one model over the others. 

\subsection{An interacting consequence}
\begin{figure}[!hb]
\centering
\includegraphics[width=0.49\textwidth]{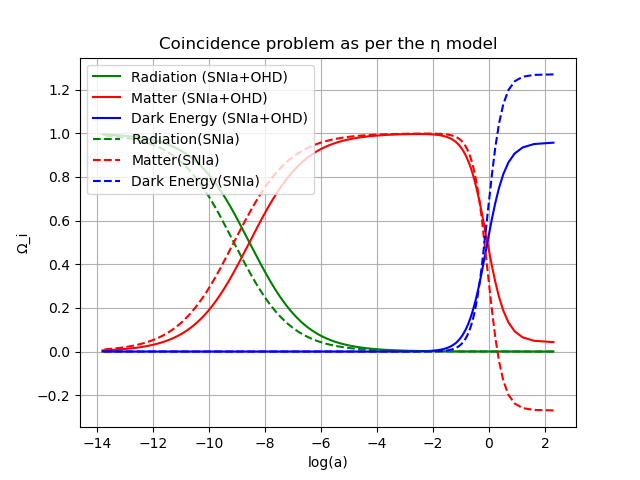}
\includegraphics[width=0.49\textwidth]{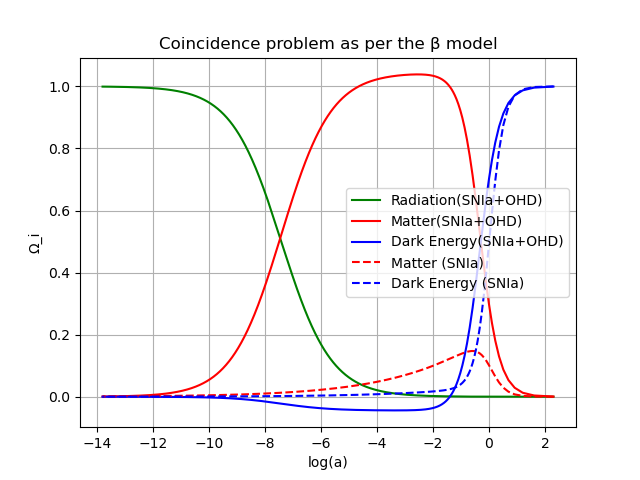}
\caption{\label{fig-sidebyside} (\emph{left})Scaled evolution of densities for the $\eta$ model. (\emph{right}) Scaled evolution of densities for the $\beta$ model. Densities follow the same colour codes, where a = 0, represent the modern universe as seen today. Beyond this is considered into the future and behind the past.}
\end{figure}
Referring to the left-hand side of Figure 2, the $\eta$ model offers the most viable solution to the coincidence problem. The SNIa data, represented by the dotted line, suggests a negative interaction (M $\rightarrow$ DE), leading to unphysical negative energy densities. However, the OHD + SNIa data provides a more consistent outcome, where the dark energy density and dark matter density remain of the same magnitude in future times. Similarly, in the $\beta$ model, both the $M \rightarrow DE$ and $DE \rightarrow M$ interactions exacerbate the coincidence problem. The SNIa data produced unphysical negative energy densities, further worsening the issue. Additionally, the OHD + SNIa data does not produce a clear matter-dominated epoch, which is crucial for large-scale structure formation \cite{10.1093/mnras/sty879}. This could represent the first significant flaw of the $\beta$ model.
\section{Conclusions}

The MCMC analysis of the $\beta$ and $\eta$ models produced a Hubble parameter of $72.158 \pm 0.275$ and $72.371 \pm 0.355$. A correction supplied by the OHD data had no significant adjustment on the Hubble constant, but had a rather significant impact on the coupling. The correction changed both models from relatively larger $M\rightarrow DE$ interaction to a small $DE\rightarrow M$ interaction.
\\ \\
The results of the MCMC analysis clearly indicate that the Pantheon dataset favours an interaction $M\rightarrow DE$ interaction in both the $\eta$ and $\beta$ model, particularly evident from the inferred densities. Furthermore, the coincidence problem was worsened by the $M\rightarrow DE$  interaction as other papers suggested. When OHD data was added to the Pantheon data set, a $DE\rightarrow M$ was found but no significant conclusion could be drawn up due to the large discrepancy between datasets and claims of biased results due to systematic errors in age determinations \cite{M_Moresco_2012}. Clearly, the $\beta$ model provides small matter dominated epoch, which in turn raises questions about early universe galaxy formation. The $\eta$ model effectively addressed the coincidence problem but requires an interaction from $DE\rightarrow M$, which could not be definitively established.
\\ \\
The $\eta$ model suggests that an interaction between dark energy (DE) and dark matter (M) can alleviate the coincidence problem. This is because the interaction allows the densities of dark energy and dark matter to evolve in such a way that they remain comparable in magnitude over time, without leading to unphysical negative energy densities. This gradual convergence of the two densities helps resolve the issue of their seemingly fine-tuned values at present times, which is a key feature of the coincidence problem. The $\eta$ model shows great promise and should be considered for further investigation focused on structure growth would be valuable to critically assess the model's strengths and limitations.

%Bibliography
%make sure you fill the .bib properly, e.g. edit "iopart-num.bib" and add all your references in there.
\section*{References}
\bibliographystyle{iopart-num}
\bibliography{323_RG_Rugg.bib}

\end{document}